# Generalized Isoscaling of Isotopic Distributions

R. Shomin, M.B. Tsang, O. Bjarki, C.K. Gelbke, G.J. Kunde[1], R.C. Lemmon[2], W.G. Lynch, D. Magestro, R. Popescu[3], A.M. Vandermolen, G. Verde, G.D. Westfall, H.F. Xi

*Department of Physics and Astronomy and National Superconducting Cyclotron Laboratory, Michigan State University, East Lansing, MI 48824, USA,*

W. A. Friedman

*Department of Physics, University of Wisconsin, Madison, WI 53706, USA*

G. Imme, V. Maddalena, C. Nociforo, G. Raciti, G. Riccobene, F.P. Romano, A. Saija, C. Sfienti,

*Dipartimento di Fisica Università and I.N.F.N. Sezione and Laboratorio Nazionale del SUD - Catania, I95127, Catania, Italy*

S. Fritz, C. Groß, T. Odeh, C. Schwarz,

*Gesellschaft für Schwerionenforschung, D-64220 Darmstadt, Germany,*

A. Nadasen, D. Sisan, K.A.G. Rao

*Department of Natural Sciences, University of Michigan, Dearborn, MI 48128, USA*

Abstract

Generalized isoscaling relationships are proposed that may permit one to relate the isotopic distributions of systems that may not be at the same temperature. The proposed relationships are applied to multifragmentation excitation functions for central Kr+Nb and Ar+Sc collisions.

---

[1] Present address: Yale University, New Haven CT 06520, U.S.A.
[2] Present address: CCLRC, Daresbury Laboratory, Daresbury, Cheshire WA44AD, U.K.
[3] Present address: Brookhaven National Laboratory, Upton, NY 11973, U.S.A.

At incident energies in excess of about E/A=30 MeV, a rapid collective expansion of the combined system may occur during the later stages of a central collision between heavy nuclei [1,2]. At densities less than about 1/3 saturation density such systems disassemble into a mixture of fragments and light particles; the duration of fragment emission is of the order of 100 fm/c [3,4]. Even though the emission time is short, statistical models such as the bulk multifragmentation models which assume equilibrium at a single breakup density and temperature, have been used successfully to describe many experimental observables such as the fragment multiplicities, charge distributions, and the energy spectra of the emitted fragments [2,5-7]. These descriptions require careful, though not necessarily obvious, choices for the source size, excitation energy and collective velocity of expansion [2,6,7,8]. Many of these statistical models display a phase transition in nuclear matter with sub-saturation density [9,10]; such models have been employed to extract the caloric curve, i.e. the relationship between excitation energy and temperature for the nuclear liquid-gas phase transition [5,11-17] and to address whether finite nuclear systems may display negative heat capacities in analogy to those deduced for finite metallic clusters [18].

The success of thermal models raises the fundamental question of whether local thermal equilibrium is achieved in such collisions. It is important to note that there are problems with determining both the excitation energy [2,5,7,19] and the temperature [12-15,20-26] of multifragmenting systems. After correction for collective expansion, for example, calculated excitation energies for peripheral collisions at high energies must be further reduced by roughly 30% to reproduce experimental data and larger corrections are estimated for energetic central collisions [2,7,19,27]. Collective motion, pre-equilibrium emission and Coulomb barrier fluctuations increase significantly the temperatures deduced from kinetic energy spectra [1-3,12,19] while secondary decay modifies the temperatures deduced from excited state populations and isotope ratios [21,22]. Resolving these problems is an important priority.

Regardless of whether thermal equilibrium is actually achieved in such collisions, thermal models will remain extremely useful because they suggest a simple and transparent description of the data. Here, we focus upon the description of isotopic data where thermal models have suggested an isotopic thermometer obtained from a double ratio of isotopic yields and an isoscaling relationship obtained from a single ratio of isotopic yields. The isoscaling relationship provides a remarkably accurate way to relate data for systems of different isotopic composition but with approximately the same excitation energy per nucleon or temperature [8,28-30].

In this paper, we begin by introducing these basic observables. We then generalize the isoscaling relationship to allow comparisons of systems at different temperatures and examine the accuracy of this generalization. We use the isotopic thermometer to provide input regarding the temperature difference required by the generalized isoscaling relationship. We study the empirical comparison between the isotopic temperatures and generalized isoscaling parameters obtained for a heavier Kr+Nb and those obtained for a lighter Ar+Sc system as a function of the incident energy per nucleon.

## Isotopic thermometers and isoscaling parameterizations

Due to the ease of measuring isotope cross-sections, the most widely used experimental method to measure temperatures in the caloric curves is to determine the relative isotopic abundances of two pairs of isotopes with large binding energy differences, B [11-17,20-23,25,26]. Most experimental isotope temperature determinations have used the following expression [23]:

$$T_{iso} = \frac{B}{\ln(a \cdot R)} \qquad (1)$$

where $R = (Y(3)/Y(4))/(Y(1)/Y(2))$ is fragment yield ratio of the ground states for isotope pairs (3,4) and (1,2), $a$ is a ground state spin factor. Information on the four thermometers studied in the present work is listed in Table I.

Eq. (1) assumes that the excited systems are at statistical equilibrium and that the systems can be approximated by grand canonical ensembles i.e. finite size effects and effects of sequential decays on the isotope yields are negligible. However, the sequential decay effects can be significant and at high temperatures dependent on the contributions of very short-lived unbound resonances [16,17,22,25,31-35]. These contributions can be calculated subject to certain model dependent assumptions about the continuum contributions and determined by direct measurements of the decays of these unbound states

It has been found empirically that isotope ratios from two statistical processes, 1 and 2, with same temperature exhibit isoscaling [28,29, 36], i.e. the isotope ratios depends exponentially on the neutron number, N, and proton number, Z, of the isotope (N, Z)

$$R_{21}(N,Z)=Y_2(N,Z)/Y_1(N,Z)=C\exp(\alpha N+\beta Z) \qquad (2)$$

where α and β can be treated as empirical fitting parameters and C is the overall normalization factor. Eqs. (1) and (2) can be derived from the simple Grand Canonical model expression for the primary fragment yield for $i^{th}$ fragment in its $k^{th}$ state before secondary decay:

$$Y_{i,primary} \cup V \frac{A_i^{3/2}}{\lambda_T^3}(2J_{ik}+1)\exp[(Z_i\mu_P + N_i\mu_n + B_{ik})/T] \qquad (3)$$

where $\mu_p$ and $\mu_n$ are the proton and neutron chemical potentials, $\lambda_T = h/\sqrt{2mT}$, $B_{ik}$ and $J_{ik}$ are the binding energy and spin of the fragment in the $k^{th}$ state, and V is the free (unoccupied) volume of the system. The insertion of the ground state yields predicted by Eq. (3) into Eq. 2 results in the cancellations of binding energy terms provided the temperatures of the two reactions are equal. Similarly, the insertion of the ground state yields predicted by Eq. (3) into Eq. (1) results in the cancellation of the chemical potential terms; the spin and mass number terms contribute to the factor *a* in Eq (1).

What is measured in an experiment, however, are the secondary yields after sequential decay. Calculations of the yields of secondary fragments after sequential decay require an accurate accounting for feeding from the particle decay of highly excited heavier nuclei [25,32,33]. Such calculations are doable but are somewhat non-transparent and subject to uncertainties regarding the levels that can be excited and the structure effects that govern their decay [25,26,32,33]. To construct simple thermal expression, we adopt instead the thermal expressions in Eqs. 1 and 2 as rough empirical guides to the possible relationships between the temperature and the charge and mass distributions and explore the extent to which they can be used to describe experimental observations. A similar approach has been taken with Eq. (2) in refs. [28,29] and justified therein by statistical model calculations [30], which suggests that secondary decay corrections largely cancel when the two systems are at the same temperature. Likewise, this approach has also been taken with the isotope thermometric expression in Eq. (1); discussions of the modifications of Eq. (1) can be found in refs. [25,26,32,33].

We take this approach in order to see whether the isoscaling relationship can be extended to consider two systems at different temperatures. In general, the binding energy factors in Eq. (3) are not cancelled by the ratio in Eq. (2) if the two systems have different temperatures. However, one may try to extrapolate the isoscaling behavior to systems with different temperatures by multiplying $R_{21}(N,Z)$ by a binding energy dependent term:

$$R_{21}(N,Z)\exp(k_{21}\cdot BE(N,Z)) = C'\exp(\alpha'N+\beta'Z), \qquad (4)$$

where $k_{21} = 1/T_1 - 1/T_2$ is a temperature dependent correction factor [37]. Because the two systems are at different temperatures, the scaling relationship of Eq. (4) may be more sensitive than that of Eq. (2) to the temperature dependent secondary decay corrections to the isotopic yields. Earlier observation of isoscaling law suggests that the sequential decay correction appears to behave as a multiplicative factor to the primary yield in Eq. 3 [28]. In this case, the temperature dependent secondary decay for two different temperature is included in the normalization constant, C', in Eq. 6. In the following, we will use measured isotope ratio temperatures in Eq. (3) to test whether empirical isotope temperatures and the generalized isoscaling relationship in Eq. (4) can describe the evolution of the isotope distributions with excitation energy. We note that it might be possible to invert Eq. (4) and obtain a temperature for one system if the temperature of the other is known.

## Experimental Analyses

The experiment was performed by bombarding $^{45}$Sc targets of 3 mg/cm$^2$ areal density with $^{36}$Ar beams at E/A=50, 100, 150 MeV and $^{93}$Nb targets of 3 mg/cm$^2$ areal density with $^{86}$Kr beams at E/A=35, 50, 100, and 120 MeV at the National Superconducting Cyclotron Laboratory at Michigan State University (MSU) [16]. Impact parameters were selected by assuming that the average multiplicity of identified charged particles detected at polar angles of 7° - 157° with 215 plastic ΔE-E phoswich of the MSU 4π array decreases monotonically with impact parameter [38,39]. Central collisions were defined by the requirement that the multiplicity N of identified charged particles lies within the highest 20% of the multiplicity distribution for N≥3. If $b_{max}$ denotes the impact parameter corresponding to <N> ≈ 3, this centrality requirement corresponds to values of the reduced impact parameter $\hat{b} = b/b_{max}$ of $\hat{b}$<0.45. The same reduced impact parameter criterion is used for all the reactions studied in this article.

We replaced two hexagonal modules of the MSU 4π array, located at 37° and 79° by 96 telescopes that covered approximate polar and azimuth angular ranges of 43° and 40°, respectively, in the laboratory [16]. To provide good coverage for light charged particles emitted at center-of-mass angles of 90°, where contributions from the decay of projectile-like and target-like fragments are minimal, the central angle of the hodoscope was placed at 47.9°, 42.6°, and 40.6° at incident energies of E/A=50, 100, 150 MeV, respectively. Each of these telescopes subtended a solid angle of 1.83 msr and consisted of a 300 um thick silicon detector followed by 6 cm thick CsI(Tl) scintillation detector. The centers of neighboring telescopes were separated by relative angles of 3.3°.

The silicon detectors of these telescopes were calibrated to an accuracy of 2% using a precision pulser and alpha particles emitted from a $^{212}$Po source. The CsI(Tl) scintillators were calibrated to an accuracy of 3% with recoil protons elastically scattered from a $CH_2$ target by $^{86}$Kr ions at E/A=35 MeV and $^4$He ions at E/A=22 and 40 MeV. The accuracies (≈3%) of the overall calibrations are largely governed by the accuracy of the CsI(Tl) calibrations. With these telescopes, the spectra of isotopically resolved light particles with Z less than five were measured at angles between 20° and 70° in the laboratory frame.

The experimental apparatus at each incident energy samples somewhat different kinematic regions of the reaction. To compare similar kinematic regions for the various incident energies and target-projectile combinations and to minimize contaminations from the projectile- and target-like spectators, we have extracted the isotope yields for elements Z=1-4 at $\theta_{CM}$=80°-100° with center of mass energy threshold of 5 MeV per nucleon for all isotopes. The extracted yields lie within the acceptance of the hodoscope; they were obtained by fitting the experimental data and using the fits to predict the corresponding center of mass yields. The uncertainties in the fitted yields were determined by varying the fits. The energy thresholds are necessary because the experimental set up does not provide measurements at very forward angles. The low energy thresholds also minimize contributions from the evaporation of the residues at the lowest incident energies.

Fig. 1 shows the excitation function of the isotopic temperature measurements for the two systems, Kr+Nb (left panel) and Ar+Sc reactions (right panel). Different symbols represent different thermometers as specified in the left panel. The lines are drawn to guide the eyes. The increase of the ratio temperatures with incident energy seems to be more systematic for the Kr+Nb system. At the incident energy of 100A MeV where both systems have measurements, the isotope ratio temperatures obtained in the Ar+Sc are similar to the temperatures obtained in the Kr+Nb system within the experimental uncertainties. Due to the large uncertainties, we cannot determine the dependence of isotopic temperatures on the system size [14].

Earlier studies have attributed the differences between the apparent temperatures for the different isotope pairs to the influence of the secondary decays of the heavier excited fragments formed in the early stages of the collisions. Over a moderate temperature range, one can reduce the differences between the thermometers by using the empirical relationship [15,20,21,24,]

$$1/T_o = 1/T_{app} - \ln\kappa/B \qquad (5)$$

where the values of $\ln\kappa/B$ depends on specific isotope pairs used. Following this empirical approach, we applied the correction factors from Refs. [20,21] as listed in Table I to all the isotopic temperatures in Fig.1. The resulting values for $T_0$, shown in Fig. 2, display an increase with incident energy, but the variations between different thermometers are much smaller than for the measured temperature $T_{app}$. The variations in the corrected temperatures, $T_0$, are larger for the Ar+Sc system (right panel) than for the Kr+Nb system (left panel). Unfortunately, the uncertainties in the extracted values are too large to draw definitive conclusions.

To examine whether a generalized isoscaling can be applied to these reactions, we construct the isotope ratios, $R_{21}(N,Z)$ from measurements on the same system at two different incident energies. The top panels of Fig. 3 show the isotope ratios measured in Kr+Nb collisions and the top panels in Fig. 4 show the isotope ratios measured in Ar+Sc collisions for Z=1, (open circles), Z=2 (closed circles), Z=3 (open squares), and Z=4 (closed squares) isotopes and different combinations of incident energies. The different incident energies involved in each ratio are labeled in each panel; e.g. the notation "70/35" in the upper left panel in Fig. 3 denotes the ratio of isotopic yields measured at E/A=75 MeV in Kr+Nb collisions to the corresponding yields measured at E/A=35 MeV. For simplicity, we adopt the convention that isotope yields from the higher energy collision are placed in the numerator. Clearly, the raw isotopic ratios in the upper panels of these figures don't show any systematic trend. Instead, the ratios fluctuate from isotope to isotope by a factor of two.

To determine whether these fluctuations are consistent with the binding energy term that results from a difference between the temperatures $T_1$ and $T_2$ for the two reactions measured at incident energies of $E_1$ and $E_2$, we compensated approximately for the temperature difference using Eq. 2. For $k_{1,2}$, we used the average value $\langle k_{app} \rangle$ where:

$$\langle k_{app} \rangle = \langle 1/T_{1,app} - 1/T_{2,app} \rangle \tag{6}$$

where $T_{1,app}$ and $T_{2,app}$ are the measured isotopic temperature for a specific isotopic thermometer plotted in Fig. 1. The average is taken over all of the isotopic thermometers. (If To values from Figure 2 are used, same results are obtained.) These corresponding mean values $\langle k_{app} \rangle$ given in the fourth column of Table 2 and used as labels for the lower panels of Fig. 3 where the adjusted isotope ratios, $R_{21}(N,Z)\exp(\langle k_{app} \rangle \cdot BE(N,Z))$ are shown as the open and closed points. The degree to which this procedure removes the fluctuations in isotopic ratios in Fig. 3 is remarkable. Alternatively, one can extract the k values by fitting the $R_{21}$ data in the top panels of Fig. 3 with Eq. (3). These best fit values, given in the column in Table 2 labeled $k_{fit}$, are statistically consistent with the mean values of $\langle k_{app} \rangle$. The values for α' and β' that describe the dependence in Eqs. 3 and 4 upon neutron and proton number are also given in the table.

When one performs the same procedure for Ar+Sc collisions, mean values of $\langle k_{app} \rangle$ = -0.039 and -0.028 are obtained for the pairs of incident energies involved in the left and right panels of Fig. 4. If the values for $\langle k_{app} \rangle$ are used to adjust the isotope ratios as shown in the lower panels of Fig. 4, the adjusted ratios scaled to a small range of values and seems to obey Eq 4 with the value of parameter β close to zero. Due to large uncertainties in the fit, we cannot determine if the $\langle k_{app} \rangle$ values obtained are consistent with the corresponding best fit values for $k_{21}$, see Table II.

Since the temperatures of the Kr+Nb system and Ar+Sc at E/A=100 MeV are similar, we compare the isotope yield ratios for the Ar+Sc system and Kr+Nb system at this energy. The raw isotope ratios $R_{21}$ is shown in the left panel of Figure 5. All the ratios seem to lie in a narrow range. If we try to fit all the isotope ratios using Eq. 4, the best fit $\langle k_{app} \rangle$ value is –0.019. The corrected ratios with the best fit lines are shown in the right panel of Figure 5. The best fit value is consistent with the calculated value of $k_{21}$ =1/$T_{app}$(Kr+Nb)-1/ $T_{app}$(Ar+Sc)= -0.010±0.009 MeV. This suggests that the Ar+Sc system is not too far from equilibrium as the Kr+Nb system especially if slightly higher temperature is assumed for the lighter system. The slightly higher temperature for Ar+Sc system is consistent with previous studies of limiting temperature on source size [14].

Dynamical stochastic mean field calculations suggest that the yields of excited fragments produced by dynamical models are not as consistent with isoscaling relations as the final yields after secondary decay [40]. Thus, the consistency of the Kr+Nb data with the generalized scaling relationships could be due to a higher degree of equilibration in the heavier system or to a greater abundance of heavier fragments that sequentially decay to the observed ones. The quality of the data in the Ar+Sc reaction is not sufficient to draw such a conclusion. In any case improved measurements of systems with different sizes and with larger range of isotopes would be useful to establish the validity of generalized isoscaling more clearly.

In summary, evidence for the validity of generalized isoscaling relation that allows one to relate systems of different isotopic composition and at different excitation energies is observed. The accuracy of these generalized scaling relationships indicate that equilibrium concepts may provide a more reasonable approximation to the final state for the collisions of heavy systems collisions. Even though the generalized isoscaling is a necessary but not a sufficient condition for the existence of equilibrium, it may be interesting to use this observable to investigate if the equilibrium conditions are established in the measurements of the caloric curves especially in the plateau region where temperatures are nearly constant.

**Figures Captions**:

Fig. 1: Apparent temperatures (Eq. 1) extracted from the four isotope ratio thermometers listed in Table 1 as a function of incident energy for the Kr+Nb system (left panel) and Ar+Sc system (right panel). The lines are drawn to guide the eye.

Fig. 2. Corrected temperatures, $T_o$ (Eq. 5), for the Kr+Nb system (left panel) and Ar+Sc system (right panel).

Fig. 3. Top panel: Relative isotope yield ratios for Z=1 (open circles), Z=2 (closed circles), Z=3 (open squares), Z=4 (closed squares), for the Kr+Nb reactions. The ratios are obtained using the isotope yields from two different incident energies. The energies involved are labeled in the top panels as $E_2/E_1$. See text for details. Bottom Panels: Relative isotope yield ratios corrected for temperature differences using Eq. 3 are shown as the open and closed points. The lines are drawn to guide the eye.

Fig. 4: Same as Fig. 3 for the Ar+Sc reactions. The lines are drawn to guide the eye.

Fig. 5: Comparison of isotope yields for the Kr+Nb and Ar+Sc reactions for incident energy of 100 MeV. Left panel: raw yield ratios for Z=1 (open circles), Z=2 (closed circles), Z=3 (open squares), Z=4 (closed squares). Right panel, corrected ratios using the best fit parameter k=-0.019.

**Table I**: Four thermometers and their relevant parameters used in this article

| Thermometer | Isotope Ratio | $a$ | $B$ (MeV) | $(\ln\kappa/B)$ (MeV$^{-1}$) |
|---|---|---|---|---|
| $T_{He}^{6,7Li}$ | ($^{6,7}$Li, $^{3,4}$He) | 2.18 | 13.32 | -0.0051 |
| $T_{He}^{2,3H}$ | ($^{2,3}$H, $^{3,4}$He) | 1.59 | 14.29 | 0.0097 |

| $T_{He}^{1,2}{}_H$ | ($^{1,2}$H, $^{3,4}$He) | 5.60 | 18.4 | 0.0496 |
| $T_{He}^{7,8}{}_{Li}$ | ($^{7,8}$Li, $^{3,4}$He) | 1.98 | 18.54 | 0.0265 |

**Table II.** Generalized scaling parameters. The third and fourth columns list the values for $\langle k_{app} \rangle$ obtained from averaging the data in Fig. 1 and from the best fit, respectively. The values for α' and β' are weighted average of the scaling parameters from fits using $\langle k_{app} \rangle$ and $k_{fit}$ to describe the temperature dependence.

| Collision | $E_2$(MeV)/$E_1$(MeV) | $\langle k_{app} \rangle$ | $k_{fit}$ | α' | β' |
|---|---|---|---|---|---|
| Kr+Nb | 70/35 | -0.049±0.005 | -0.040±0.005 | 0.3489 | 0.4034 |
| Kr+Nb | 100/70 | -0.024±0.005 | -0.024±0.004 | 0.0885 | 0.2073 |
| Kr+Nb | 120/70 | -0.034±0.005 | -0.028±0.003 | 0.1561 | 0.2340 |
| Ar+Sc | 100/50 | -0.045±0.005 | -0.028±0.004 | 0.2962 | 0.1461 |
| Ar+Sc | 150/100 | -0.025±0.005 | -0.025±0.003 | 0.2303 | -0.051 |

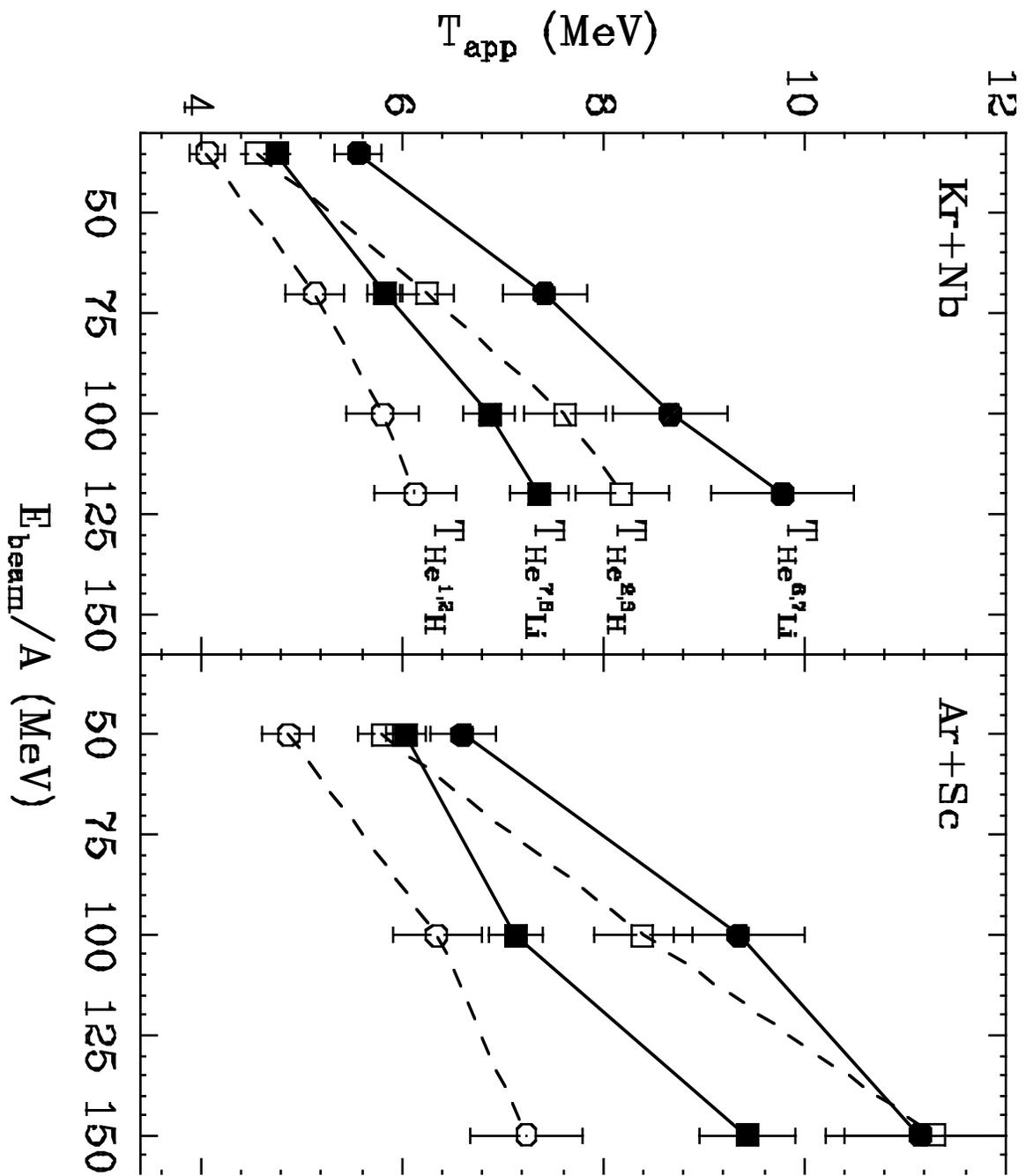

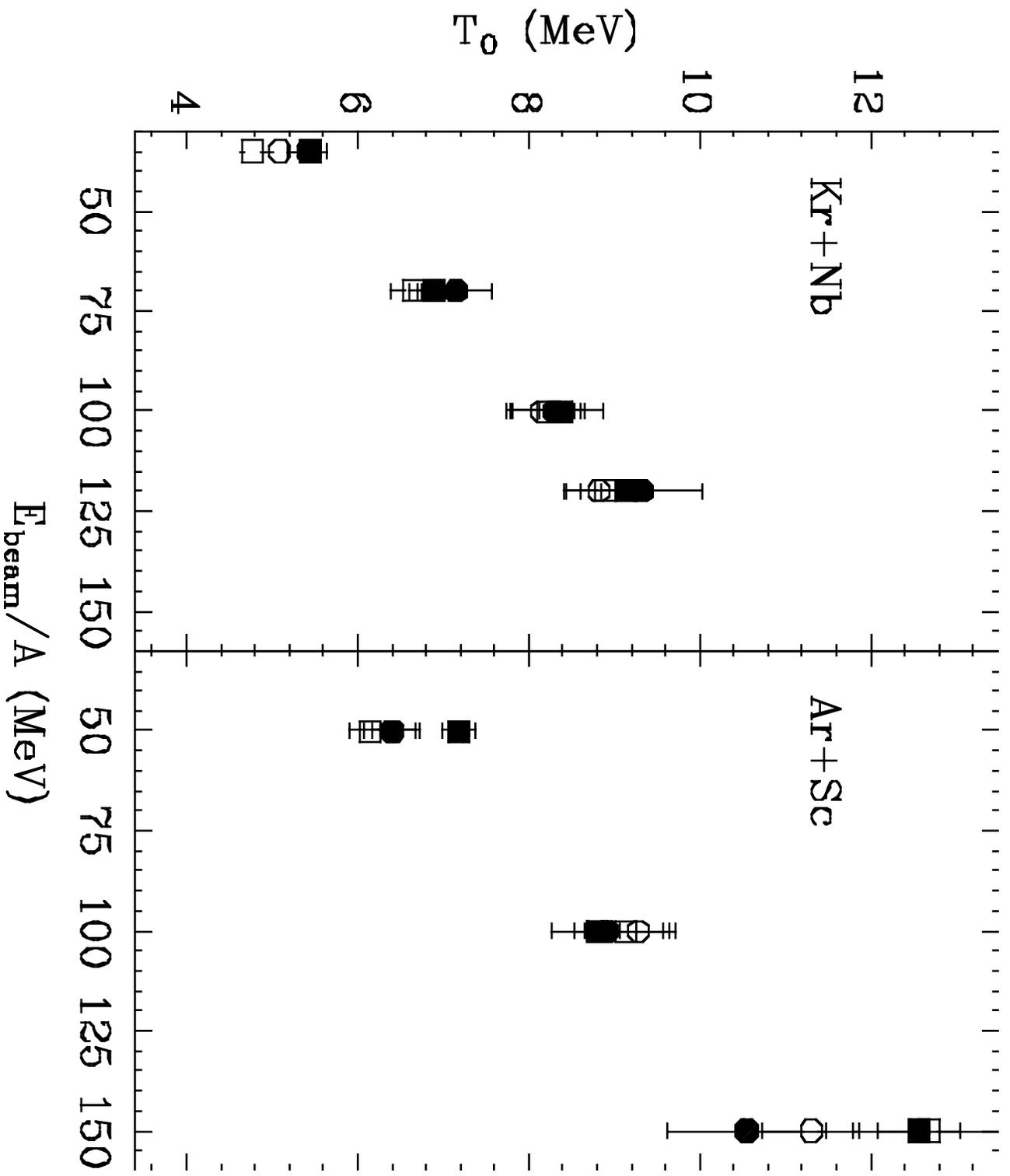

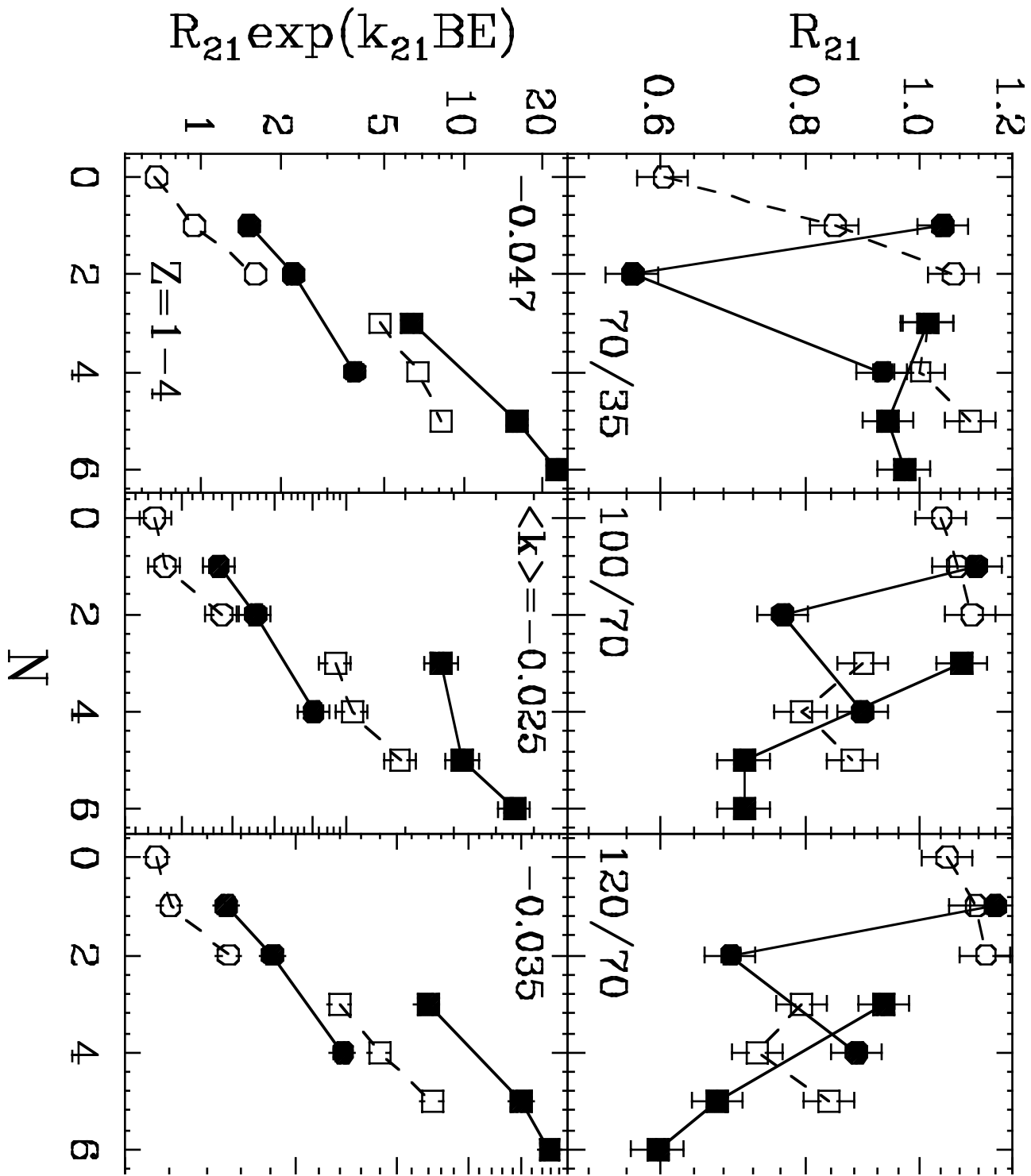

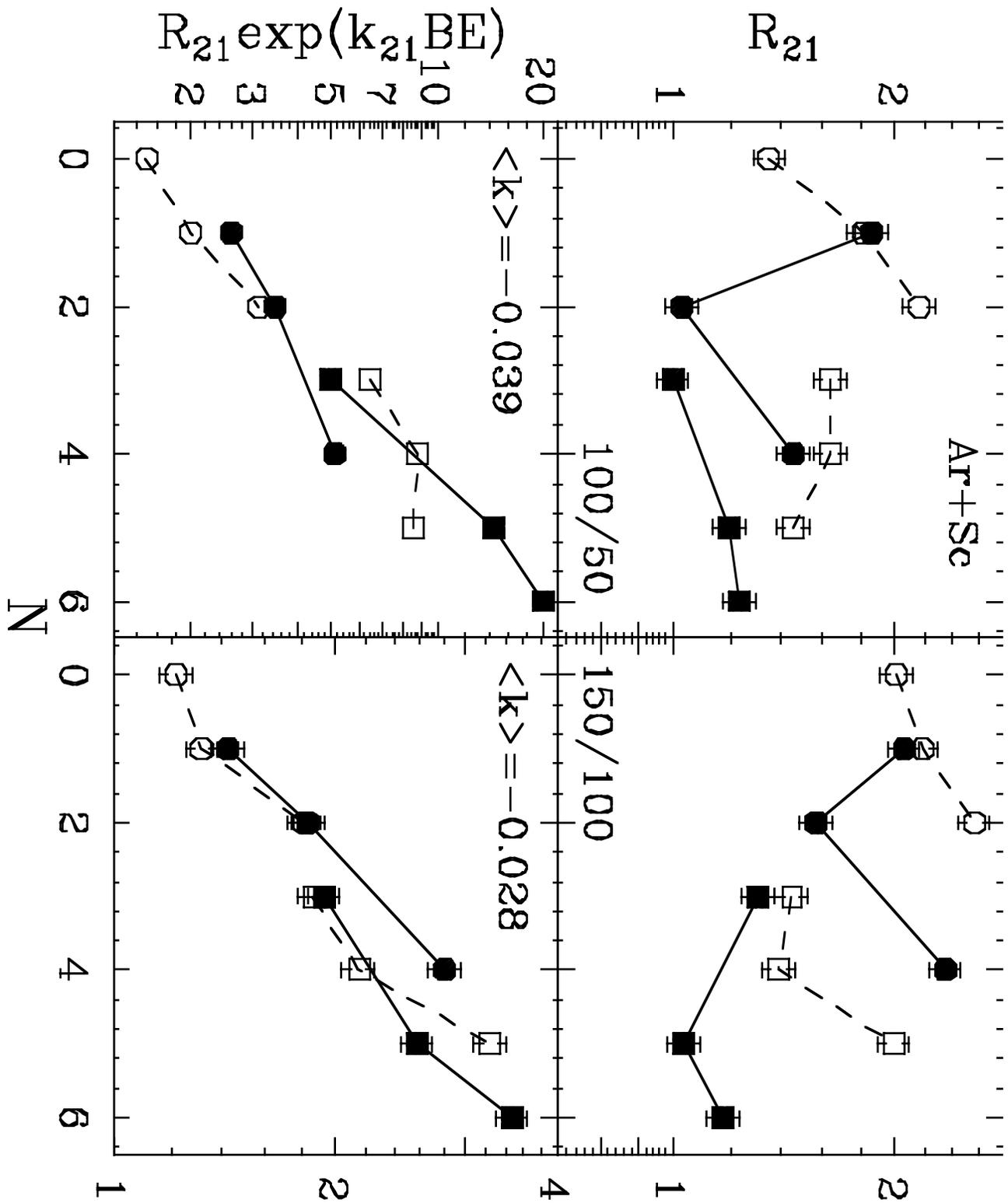

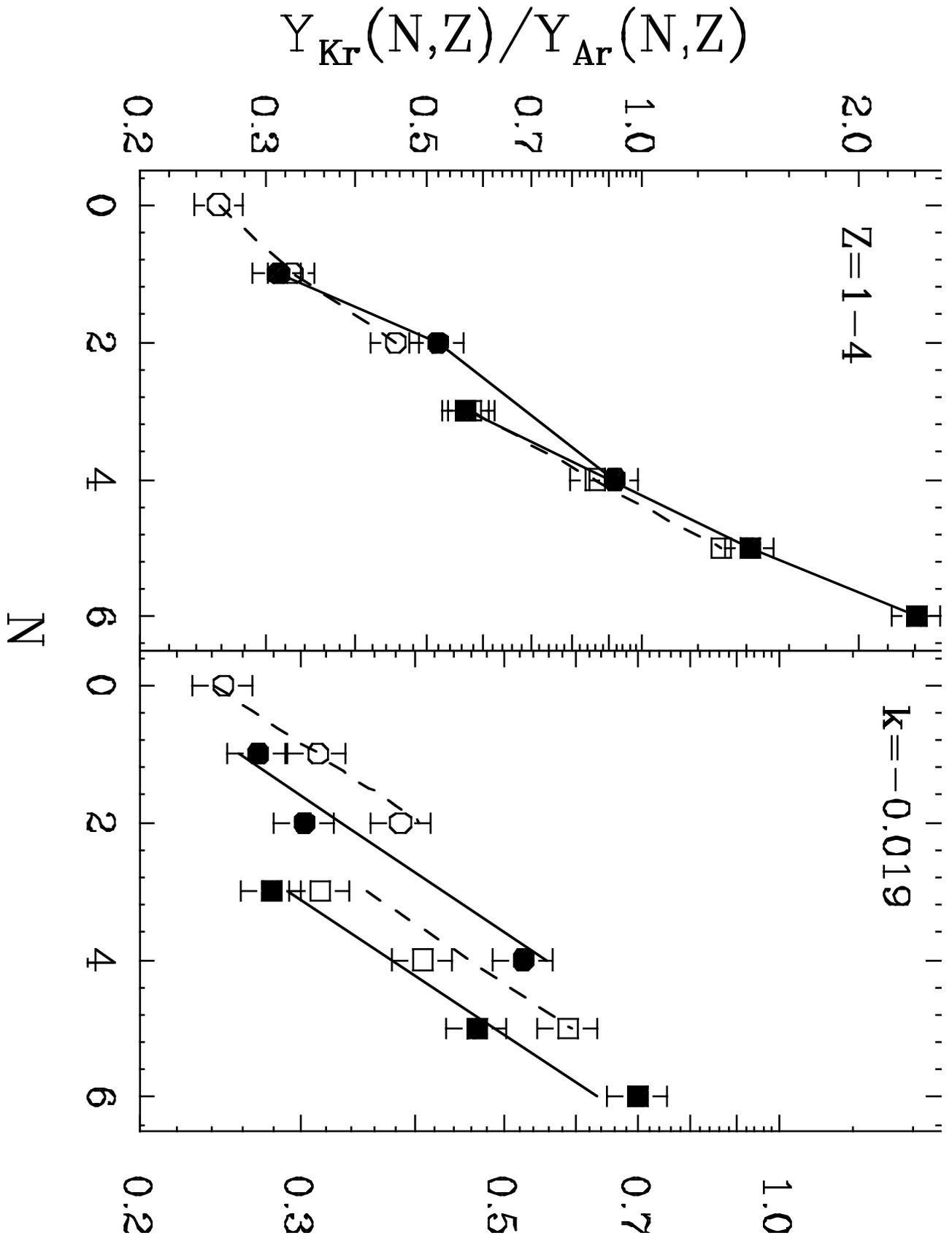

20